\documentclass[reprint,prd,twocolumn,superscriptaddress]{revtex4-1}

\usepackage{color}
\usepackage{amsmath}
\usepackage{graphicx}
\newcommand{\evenspacing}{\vphantom{\bigg[}}
\begin{document}
\hfill TTP19-039,  P3H-19-045
\title{Top Quark Mass Effects in Next-To-Next-To-Next-To-Leading Order Higgs
  Boson Production: Virtual Corrections}
\author{Joshua Davies}
\email{joshua.davies@kit.edu}
\author{Florian Herren}
\email{florian.herren@kit.edu}
\author{Matthias Steinhauser}
\email{matthias.steinhauser@kit.edu}
\affiliation{Institut f\"ur Theoretische Teilchenphysik,
  Karlsruhe Institute of Technology (KIT), 76128 Karlsruhe, Germany}
\begin{abstract}
  We compute four-loop corrections to the Higgs boson gluon vertex, including
  finite top quark mass effects. Analytic results are presented which serve as
  a building block for the next-to-next-to-next-to-leading order corrections
  to Higgs boson production at the Large Hadron Collider at CERN.
\end{abstract}
\maketitle

{\bf Introduction.}  The precise measurement of the Standard Model Higgs boson
properties is a major focus of the physics program of the Large Hadron
Collider (LHC) at CERN. A crucial quantity in this context is the total cross
section for Higgs boson production in proton-proton collisions. There are
several mechanisms which contribute to the cross section. The largest contribution
is from the gluon-gluon fusion process~\cite{deFlorian:2016spz} despite it being
loop induced. It is thus important to have precise control over its higher-order
quantum corrections.

Leading order (LO) corrections to the process $gg\to H$ were considered
40~years ago~\cite{Georgi:1977gs} and next-to-leading order (NLO) corrections
were computed at the beginning of the nineties, first in the infinite--top
quark mass approximation~\cite{Djouadi:1991tka,Dawson:1990zj} and shortly
after exactly in $m_t$~\cite{Spira:1995rr}.  (The analytic two-loop
virtual corrections are known
from~\cite{Harlander:2005rq,Anastasiou:2006hc,Aglietti:2006tp}.)  About ten
years later the first next-to-next-to-leading order (NNLO) results became
available~\cite{Harlander:2002wh,Anastasiou:2002yz,Ravindran:2003um} in an
effective-theory framework where the top quark is integrated out. It took a
further ten years to compute corrections in $m_H/m_t$ and estimate the
finite-$m_t$ effects.
Several expansion terms of the Higgs-gluon form factor were computed
in~\cite{Harlander:2009bw,Pak:2009bx} and subsequently an approximation method was
developed to treat also the real-radiation
contribution~\cite{Harlander:2009mq,Pak:2009dg,Harlander:2009my,Pak:2011hs}.
Recently three-loop corrections to the Higgs-gluon vertex with a massive quark
loop have been obtained by combining information from the large-$m_t$
and threshold expansions with the help of a conformal mapping and a Pad\'e
approximation~\cite{Davies:2019nhm}. For the subset of three-loop
diagrams which contain a closed light-quark loop, even analytic results are
available~\cite{Harlander:2019ioe}.
The next-to-next-to-next-to-leading order
(N$^3$LO) corrections to $gg\to H$ within the effective-theory approach have
been computed in Refs.~\cite{Anastasiou:2016cez,Mistlberger:2018etf}.  At this
order in perturbation theory no finite-$m_t$ corrections are available.  In
this letter we provide the first step to close this gap and compute the
finite-$m_t$ effects of the virtual N$^3$LO corrections. Note that the $gg\to H$
vertex diagrams only depend on the Higgs boson and top quark masses. Thus, it is
promising to consider an expansion for large $m_t$ which is expected to show
good convergence properties since the expansion parameter
$\rho = m_H^2/m_t^2 \approx 0.5$ is sufficiently small.  Both at NLO and NNLO it has
been shown that three expansion terms are adequate to obtain
results which, from the phenomenological point of view, are equivalent to an
exact calculation~\cite{Harlander:2009bw,Pak:2009bx}.

In this work we concentrate on the numerically dominant contributions from
diagrams in which the Higgs couples to a top quark loop. Note that although the
Yukawa coupling is small, diagrams in which the Higgs couples to a bottom quark loop
are parametrically enhanced by large logarithms. For example the LO contribution
is proportional to $m_b^2/m_H^2 \log^2(m_b/m_H)$ and thus bottom quark corrections
should be included at lower orders.

\smallskip

\begin{figure}[t]
  \includegraphics[width=0.45\textwidth]{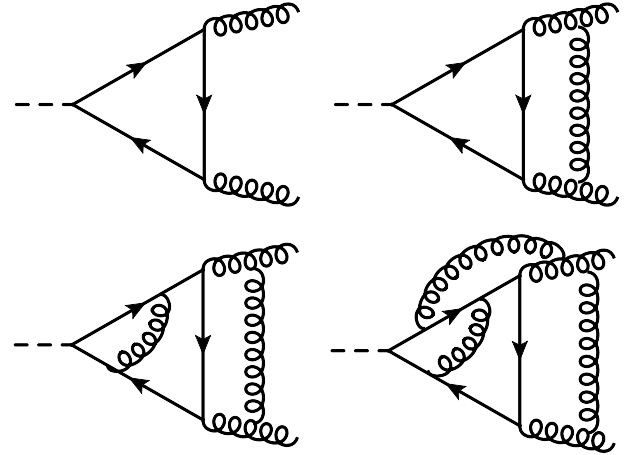}
  \caption{\label{fig::diags}Sample Feynman diagrams contributing to the process
    $gg\to H$ at LO, NLO, NNLO and N$^3$LO. The solid and curly lines
    represent top quarks and gluons, respectively. The external dashed line
    stands for the Higgs boson.}
\end{figure}

{\bf Calculation.}
The LO contribution to Higgs boson production in
gluon fusion is mediated by the one-loop diagram shown in
Fig.~\ref{fig::diags}. Correspondingly N$^k$LO corrections are
obtained from $(k+1)$-loop vertex corrections also shown in
Fig.~\ref{fig::diags}. In this letter we compute the four-loop
corrections. 

The amplitude for $gg\to H$ can be parameterized as
\begin{eqnarray}
  {\cal A} &=& {\cal A}_0 \: h(\rho) \: \delta^{ab} \left( q_1\!\cdot\! q_2\:g^{\mu\nu} - q_1^\nu q_2^\mu \right)
               \,,
               \label{eq::A}
\end{eqnarray}
where $q_1$ and $q_2$ are the momenta of the external gluons with polarization
vectors $\varepsilon^\mu(q_1)$ and $\varepsilon^\nu(q_2)$, respectively, and
$a$ and $b$ are adjoint colour indices.
${\cal A}_0$ consists of various coupling factors and mass parameters and is
given by
\begin{eqnarray}
  {\cal A}_0 &=& T_F \frac{2\alpha_s(\mu)}{3v\pi}
                 \,,
\end{eqnarray}
where $T_F=1/2$, $v$ is the vacuum expectation value and
$\alpha_s\equiv \alpha_s^{(5)}$ is defined with five active quark flavours.
The function $h(\rho)$ with $\rho=m_H^2/m_t^2$ contains the top quark mass
dependence. It has the following expansion in the strong coupling constant
\begin{eqnarray}
  h(\rho) &=& \sum_{n\ge1} \left(\frac{\alpha_s(\mu)}{\pi}\right)^{n-1}
  h^{(n)}(\rho)
  \,,
\end{eqnarray}
where the leading term is given by 
\begin{eqnarray}
  h^{(1)}(\rho) &=& \left(\frac{\mu^2}{m_t^2}\right)^{\epsilon}
                    {\Gamma(1+\epsilon)} \Bigg(1 
                    + \frac{7+7\epsilon}{120}\rho
                    \nonumber\\&&\mbox{}
                    + \frac{2+3\epsilon+\epsilon^2}{336}\rho^2
                    + {\cal O}(\rho^3)
                    \Bigg)
                    \,,
\end{eqnarray}
with $\epsilon=(4-d)/2$, where $d$ is the space-time dimension.

We apply projectors which independently project on the pre-factors of
$g^{\mu\nu}$ and $q_1^\nu q_2^\mu$ [cf. Eq.~(\ref{eq::A})] and treat the
corresponding expressions independently. In the course of the calculation the
intermediate expressions are different, however, the final results are equal
(up to an overall sign) which is a welcome check for our calculation.

We apply an asymptotic expansion~\cite{Smirnov:2013} in the limit
$m_t^2 \gg q_1\!\cdot\! q_2 = m_H^2/2$. This decomposes each Feynman diagram into
a number of so-called ``hard subgraphs'' which have to be expanded in their
external momenta. Afterwards they appear as an effective vertex in the
``co-subgraph'' which is obtained from the original Feynman diagram
after contracting all lines present in the hard subgraph to a point.  From the
technical point of view this leads to massive vacuum integrals up to four-loop
order and massless vertex integrals, up to three-loop order, where only one external leg is
off-shell. Both types of integrals have been studied in the
literature~\cite{Laporta:2002pg,Schroder:2005va,Chetyrkin:2006dh,Lee:2010hs,Baikov:2009bg,Heinrich:2009be,Lee:2010ik,Gehrmann:2010ue,Gehrmann:2010tu}.

We organize our calculation such that the vacuum integrals are computed
first. This requires that we solve tensor integrals
since the integrand in general contains scalar products between the loop momenta of
the vacuum integral and $q_1$ or $q_2$ or the loop momenta of the subsequent
massless integration. Up to two-loop order there are general algorithms which
treat tensor integrals of arbitrary rank~\cite{Chetyrkin:1993rv}. At three and four loops
we have implemented tensor integrals up to rank eight which is sufficient to
obtain expansion terms up to $\rho^2$.

The application of the asymptotic expansion leads to a separation of the
scales, at the price that the number of integrals to be computed is
increased drastically; we have to consider around 40 million
three- and four-loop vacuum integrals and 1 million one- to three-loop massless
form-factor integrals.
Because of the expansions in external momenta, many of the
propagators are raised to relatively high powers. Similarly, the massless vertex
integrations involve integrals with high powers of numerators.

We perform the reductions to master integrals 
with the help of {\tt FIRE}~\cite{Smirnov:2019qkx} and use
symmetry relations from {\tt LiteRed}~\cite{Lee:2013mka}.  The combined size of the
integral tables (as FORM Tablebases) is about 25~GB. All master integrals are
available in analytic form, both for the vacuum integrals~\cite{Lee:2010hs}
and the massless vertices~\cite{Lee:2010ik}.

The diagrams are computed by TFORM~\cite{Ruijl:2017dtg} jobs, each using 4 workers and requiring
20GB of memory. The total wall-time required by these jobs to compute the $\rho^0$,
$\rho^1$ and $\rho^2$ terms of the expansion is approximately 6, 50 and 860 days
respectively.

The renormalization of the ultra-violet (UV) poles is straightforward.
We first renormalize $\alpha_s$ and $m_t$ in the $\overline{\rm MS}$
scheme and the external gluon fields in the on-shell scheme.
We then transform the $\overline{\rm MS}$ mass to the on-shell scheme
and decouple the top quark from the running of $\alpha_s$; our
final result is expressed in terms of $\alpha_s^{(5)}$.
Note that one has to carefully include higher-order $\epsilon$ terms
in the on-shell counterterms and the decoupling relations since our
final result still contains infra-red poles.

There are several checks of our final result. First, we project on both
structures present in Eq.~(\ref{eq::A}) and check that they are equal up to a
global sign. Then, we construct the leading ($\rho^0$) term independently with
the help of the effective-theory approach, i.e., we use the effective
gluon-Higgs coupling up to four
loops~\cite{Chetyrkin:1997un,Chetyrkin:2005ia,Schroder:2005hy,Gerlach:2018hen}
and the form factor results from the
literature~\cite{Baikov:2009bg,Gehrmann:2010ue} to obtain the amplitude
${\cal A}$.  Furthermore, the remaining poles after UV renormalization agree
with the predictions provided, e.g., in
Ref.~\cite{Becher:2009cu,Gehrmann:2010ue}. Our three-loop results for the top
quark mass corrections to the form factor agree with
Refs.~\cite{Harlander:2009bw,Pak:2009bx}.

When discussing the structure of the final result it is advantageous to
extract the LO contribution and define
\begin{eqnarray}
  F &=& \frac{h(\rho)}{h^{(1)}(\rho)} \,\,=\,\, 1 + {\cal O}(\alpha_s)
        \,,
\end{eqnarray}
where an expansion in $\rho$ and in $\epsilon$ on the right-hand side is
understood.  Next, we consider $\log(F)$ since for this quantity there are
simple predictions for the remaining infra-red
poles~\cite{Becher:2009cu,Gehrmann:2010ue}. We expect that the pole
part of $\log(F)$ has no expansion in $\rho$ and that the poles are given by
the massless three-loop Higgs-gluon form factor.  Our explicit results confirm
these expectations. We thus define
\begin{eqnarray}
  \log(F) &=& \log(F)_{\rm poles} + \log(F)_{\rm finite}
              \,,
\end{eqnarray}
and find that $\log(F)_{\rm poles}$ reproduces the results given in
Ref.~\cite{Gehrmann:2010ue}.
$\log(F)_{\rm finite}$ has an expansion in $\rho$ which we
discuss in the remainder of this letter.

\smallskip

{\bf Results.} We perform our calculation using general colour structures for
the gauge group $SU(N)$. In order to present a compact expression we specify
the colour factors in the following to their numerical values for QCD. General
expressions, for both the $\overline{\mbox{MS}}$ and on-shell top quark mass,
can be found in the supplementary material~\cite{progdata}. For $\mu^2=m_t^2$
we have for $\log(F)_{\rm finite}$ in the on-shell scheme
\begin{widetext}
\begin{align}
\mathrm{log}&\left(F\right)_{\mathrm{finite}} =
\nonumber\\&\evenspacing
	+ {\frac{\alpha_s}{\pi} }\bigg(\frac{11}{4} + \frac{1}{8} \pi^{2}  - \frac{3}{4} l_{tH}^{2} + \frac{17}{135} \rho + \frac{3553}{226800} \rho^{2} \bigg)
\nonumber\\&\evenspacing
	+ \bigg(\frac{\alpha_s}{\pi}\bigg)^2\bigg[\frac{523}{108} + \frac{151}{192} \pi^{2} - \frac{499}{48} \zeta_{3}+ l_{tH}\bigg(-\frac{155}{36} + \frac{23}{48} \pi^{2} + \frac{9}{8} \zeta_{3}\bigg) + l_{tH}^{2}\bigg(-\frac{151}{48} + \frac{3}{16} \pi^{2}\bigg) - \frac{23}{48} l_{tH}^{3}
\nonumber\\&\evenspacing
	+ \rho\bigg(-\frac{15765509}{829440} + \frac{7}{1080} \pi^{2} + \frac{7}{540} \log(2) \pi^{2} + \frac{1909181}{110592} \zeta_{3} + \frac{793}{10368} l_{tH}\bigg)
\nonumber\\&\evenspacing
	+ \rho^{2}\bigg(-\frac{1013177390077}{234101145600} + \frac{857}{907200} \pi^{2} + \frac{857}{453600} \log(2) \pi^{2} + \frac{267179777}{70778880} \zeta_{3} + \frac{580759}{43545600} l_{tH}
	\bigg)\bigg]
\nonumber\\&\evenspacing
	+ \bigg(\frac{\alpha_s}{\pi}\bigg)^3\bigg[
	- \frac{18539405}{1119744} + \frac{441517}{62208} \pi^{2} - \frac{11549467}{82944} \zeta_{3} - \frac{50839}{311040} \pi^{4} - \frac{1949}{576} \pi^{2} \zeta_{3} + \frac{39307}{288} \zeta_{5}
\nonumber\\&\evenspacing
	- \frac{39}{8} \zeta_{3}^{2} - \frac{193}{7560} \pi^{6} + l_{tH} \bigg(-\frac{322955}{31104} + \frac{665}{96} \pi^{2} - \frac{3043}{144} \zeta_{3} - \frac{1801}{5760} \pi^{4} - \frac{15}{16} \pi^{2} \zeta_{3} - \frac{27}{4} \zeta_{5}\bigg)
\nonumber\\&\evenspacing
	+ l_{tH}^{2} \bigg(-\frac{58745}{3456} + \frac{1435}{576} \pi^{2} + \frac{25}{8} \zeta_{3} - \frac{33}{320} \pi^{4}\bigg) + l_{tH}^{3} \bigg(-\frac{3995}{864} + \frac{23}{96} \pi^{2}\bigg) - \frac{529}{1152} l_{tH}^{4}
\nonumber\\&\evenspacing
	+ \rho \bigg(-\frac{542872693595}{3218890752} + \frac{65743583}{55987200} \pi^{2} - \frac{4691}{9720} \log(2) \pi^{2} - \frac{6788585826089}{107296358400} \zeta_{3}
\nonumber\\&\evenspacing
	- \frac{11421210133}{1149603840} \log^4(2) + \frac{11364084757}{1149603840} \log^2(2) \pi^{2} + \frac{244657561171}{55180984320} \pi^{4} - \frac{11421210133}{47900160} \mathrm{Li}_{4}\big(1/2\big)
\nonumber\\&\evenspacing
	+ \frac{718337}{9979200} \log^5(2) - \frac{718337}{5987520} \log^3(2) \pi^{2} + \frac{46111267}{239500800} \log(2) \pi^{4} - \frac{10073}{25920} \pi^{2} \zeta_{3} - \frac{3254515597}{31933440} \zeta_{5}
\nonumber\\&\evenspacing
	- \frac{718337}{83160} \mathrm{Li}_{5}\big(1/2\big) + \frac{5327119}{11197440} l_{tH} + \frac{25639}{746496} l_{tH}^{2}\bigg)
\nonumber\\&\evenspacing
	+ \rho^{2} \bigg(-\frac{1055794361417882487061}{6681366555210547200} + \frac{4077367559}{23514624000} \pi^{2} + \frac{23157917500539717053}{117837152649216000} \zeta_{3}
\nonumber\\&\evenspacing
	- \frac{110153}{1632960} \log(2) \pi^{2} + \frac{2712037738087}{2391175987200} \log^4(2) - \frac{2729355664999}{2391175987200} \log^2(2) \pi^{2} - \frac{150868470717581}{229552894771200} \pi^{4}
\nonumber\\&\evenspacing
	+ \frac{2712037738087}{99632332800} \mathrm{Li}_{4}\big(1/2\big)  + \frac{46902913}{202176000} \log^5(2)  - \frac{46902913}{121305600} \log^3(2) \pi^{2} - \frac{8632107859}{33965568000} \log(2) \pi^{4}
\nonumber\\&\evenspacing
	- \frac{1233223}{21772800} \pi^{2} \zeta_{3} + \frac{49563452909}{4528742400} \zeta_{5}  - \frac{46902913}{1684800} \mathrm{Li}_{5}\big(1/2\big) + \frac{103150403081}{658409472000} l_{tH} + \frac{18740929}{3135283200} l_{tH}^{2}\bigg)\bigg]\,,
\label{eq::logF}
\end{align}
\end{widetext}
where $\zeta_n$ is the Riemann $\zeta$-function evaluated at $n$,
$\mbox{Li}_n$ denote polylogarithms, $l_{tH}=\log(m_t^2/m_H^2)+i\pi$ and $m_t$
is the top quark pole mass.  It is interesting to compare the finite $m_t$
corrections of $\log(F)_{\rm finite}$ for the various orders is
$\alpha_s$. The numerical evaluation of Eq.~(\ref{eq::logF}) for $m_t=173$~GeV
and $m_H=125$~GeV gives 
\begin{eqnarray}
\lefteqn{ \log(F)_{\rm finite} \approx
      a_t\left[ 
      \left( 11.07 - i 3.06 \right) + 0.07 + 0.004
      \right] }
      \nonumber\\&&\mbox{}
      +
      a_t^2\left[
      \left( 22.59 + i 13.24 \right) + \left( 1.02 + i 0.13 \right) 
                    + \left( 0.07 + i 0.01 \right)
      \right]
      \nonumber\\&&\mbox{}
      +
      a_t^3\left[ 
      \left( -73.18 + i 51.55 \right) + \left( 7.61 + i 0.85 \right) 
                    \right.\nonumber\\&&\mbox{} \left. \qquad
                    + \left( 0.70 + i 0.14 \right)
      \right]
                    \,,
\end{eqnarray} 
where $a_t = \alpha_s^{(5)}(m_t)/\pi$. For each order in $a_t$ we separately display
the $\rho^0$, $\rho^1$ and $\rho^2$ terms inside the square brackets.
One observes that the mass corrections become more important when going to
higher orders in $a_t$. At two loops the $\rho^1$ contribution only amounts to 0.6\%
of the real part of the leading term, whereas at three and four loops we have real
contributions of 5\% and 10\%, respectively.
At four-loop order the $\rho^2$ real contributions are below 1\% which justifies the
truncation of the expansion at this order; we expect that the next term is negligibly small.
In all cases the imaginary parts converge at least as well as the real parts.

If we repeat the same exercise for the $\overline{\rm MS}$ top quark mass, $\bar{m_t}$,
and set the renormalization scale to $\mu^2=\bar{m}_t^2$,
we observe smaller mass corrections; at one, two and three loops
the $\rho^1$ terms contribute 0.1\%, 2.5\% and 1.4\%, respectively, relative to the real
part of the $\rho^0$ terms.
In all cases the $\rho^2$ terms are smaller again by a factor five to ten.

For Higgs boson production the central value of the renormalization
scale is often set to $\mu^2=m_H^2/2$. Adopting the on-shell scheme for the top
quark mass, this leads to $\rho^1$ real contributions which amount to
0.5\%, 34\% and 1.8\% for two, three and four loops.
Note that the large relative correction at three loops is due to
accidental cancellations which make the leading term ($\rho^0$)
quite small at this order.

\smallskip

{\bf Conclusions.} In this letter we compute finite top quark mass effects for
the production virtual cross section of the Standard Model Higgs boson. We expand the
four-loop Higgs-gluon vertex diagrams for $m_t\gg m_H$ and show that three
expansion terms are sufficient to obtain precision results for the physical
values of $m_H$ and $m_t$. Our result is the first N$^3$LO calculation of
the Higgs production cross section which incorporates finite top quark mass
terms.  In the coming years, the LHC enters the era of precision Higgs boson
physics and quantum corrections such as those computed in this letter will
become important.

\smallskip

{\bf Acknowledgements.} This research was supported by the Deutsche
Forschungsgemeinschaft (DFG, German Research Foundation) under grant 396021762
--- TRR 257 ``Particle Physics Phenomenology after the Higgs Discovery''.
FH acknowledges the support of the DFG-funded Doctoral School KSETA.


\begin{thebibliography}{99}

%
%

\bibitem{deFlorian:2016spz}
  D.~de Florian {\it et al.} [LHC Higgs Cross Section Working Group],
  doi:10.2172/1345634, 10.23731/CYRM-2017-002
  arXiv:1610.07922 [hep-ph].

\bibitem{Georgi:1977gs}
  H.~M.~Georgi, S.~L.~Glashow, M.~E.~Machacek and D.~V.~Nanopoulos,
  Phys.\ Rev.\ Lett.\  {\bf 40} (1978) 692.
  doi:10.1103/PhysRevLett.40.692

\bibitem{Djouadi:1991tka}
  A.~Djouadi, M.~Spira and P.~M.~Zerwas,
  Phys.\ Lett.\ B {\bf 264} (1991) 440.
  doi:10.1016/0370-2693(91)90375-Z

\bibitem{Dawson:1990zj}
  S.~Dawson,
  Nucl.\ Phys.\ B {\bf 359} (1991) 283.
  doi:10.1016/0550-3213(91)90061-2

\bibitem{Spira:1995rr}
  M.~Spira, A.~Djouadi, D.~Graudenz and P.~M.~Zerwas,
  Nucl.\ Phys.\ B {\bf 453} (1995) 17
  [hep-ph/9504378].

\bibitem{Harlander:2005rq}
  R.~Harlander and P.~Kant,
  JHEP {\bf 0512} (2005) 015
  [hep-ph/0509189].

\bibitem{Anastasiou:2006hc}
  C.~Anastasiou, S.~Beerli, S.~Bucherer, A.~Daleo and Z.~Kunszt,
  JHEP {\bf 0701} (2007) 082
  [hep-ph/0611236].

\bibitem{Aglietti:2006tp}
  U.~Aglietti, R.~Bonciani, G.~Degrassi and A.~Vicini,
  JHEP {\bf 0701} (2007) 021
  [hep-ph/0611266].

\bibitem{Harlander:2002wh}
  R.~V.~Harlander and W.~B.~Kilgore,
  Phys.\ Rev.\ Lett.\  {\bf 88} (2002) 201801
  [arXiv:hep-ph/0201206].

\bibitem{Anastasiou:2002yz}
  C.~Anastasiou and K.~Melnikov,
  Nucl.\ Phys.\  B {\bf 646} (2002) 220
  [arXiv:hep-ph/0207004].

\bibitem{Ravindran:2003um}
  V.~Ravindran, J.~Smith and W.~L.~van Neerven,
  Nucl.\ Phys.\  B {\bf 665} (2003) 325
  [arXiv:hep-ph/0302135].

\bibitem{Harlander:2009bw}
  R.~V.~Harlander and K.~J.~Ozeren,
  Phys.\ Lett.\ B {\bf 679} (2009) 467
  [arXiv:0907.2997 [hep-ph]].

\bibitem{Pak:2009bx}
  A.~Pak, M.~Rogal and M.~Steinhauser,
  Phys.\ Lett.\ B {\bf 679} (2009) 473
  [arXiv:0907.2998 [hep-ph]].

\bibitem{Harlander:2009mq}
  R.~V.~Harlander and K.~J.~Ozeren,
  JHEP {\bf 0911} (2009) 088
%
  [arXiv:0909.3420 [hep-ph]].

\bibitem{Pak:2009dg}
  A.~Pak, M.~Rogal and M.~Steinhauser,
  JHEP {\bf 1002} (2010) 025
%
  [arXiv:0911.4662 [hep-ph]].

\bibitem{Harlander:2009my}
  R.~V.~Harlander, H.~Mantler, S.~Marzani and K.~J.~Ozeren,
  Eur.\ Phys.\ J.\ C {\bf 66} (2010) 359
%
  [arXiv:0912.2104 [hep-ph]].

\bibitem{Pak:2011hs}
  A.~Pak, M.~Rogal and M.~Steinhauser,
  JHEP {\bf 1109} (2011) 088
%
  [arXiv:1107.3391 [hep-ph]].

\bibitem{Davies:2019nhm}
  J.~Davies, R.~Gr\"ober, A.~Maier, T.~Rauh and M.~Steinhauser,
  Phys.\ Rev.\ D {\bf 100} (2019) no.3,  034017
%
  [arXiv:1906.00982 [hep-ph]].

\bibitem{Harlander:2019ioe}
  R.~V.~Harlander, M.~Prausa and J.~Usovitsch,
  JHEP {\bf 1910} (2019) 148
%
  [arXiv:1907.06957 [hep-ph]].

\bibitem{Anastasiou:2016cez}
  C.~Anastasiou, C.~Duhr, F.~Dulat, E.~Furlan, T.~Gehrmann, F.~Herzog,
  A.~Lazopoulos and B.~Mistlberger,
  JHEP {\bf 1605} (2016) 058
%
  [arXiv:1602.00695 [hep-ph]].

\bibitem{Mistlberger:2018etf}
  B.~Mistlberger,
  JHEP {\bf 1805} (2018) 028
%
  [arXiv:1802.00833 [hep-ph]].

\bibitem{Smirnov:2013}
  V.~A.~Smirnov, {\em Analytic tools for Feynman integrals},
  Springer Tracts Mod.\ Phys.\  {\bf 250} (2012) 1.

\bibitem{Laporta:2002pg}
  S.~Laporta,
  Phys.\ Lett.\ B {\bf 549} (2002) 115
%
  [hep-ph/0210336].

\bibitem{Schroder:2005va}
  Y.~Schroder and A.~Vuorinen,
  JHEP {\bf 0506} (2005) 051
%
  [hep-ph/0503209].

\bibitem{Chetyrkin:2006dh}
  K.~G.~Chetyrkin, M.~Faisst, C.~Sturm and M.~Tentyukov,
  Nucl.\ Phys.\ B {\bf 742} (2006) 208
%
  [hep-ph/0601165].

\bibitem{Lee:2010hs}
  R.~N.~Lee and I.~S.~Terekhov,
  JHEP {\bf 1101} (2011) 068
%
  [arXiv:1010.6117 [hep-ph]].

\bibitem{Baikov:2009bg}
  P.~A.~Baikov, K.~G.~Chetyrkin, A.~V.~Smirnov, V.~A.~Smirnov and
  M.~Steinhauser,
  Phys.\ Rev.\ Lett.\  {\bf 102} (2009) 212002
%
  [arXiv:0902.3519 [hep-ph]].

\bibitem{Heinrich:2009be}
  G.~Heinrich, T.~Huber, D.~A.~Kosower and V.~A.~Smirnov,
  Phys.\ Lett.\ B {\bf 678} (2009) 359
%
  [arXiv:0902.3512 [hep-ph]].

\bibitem{Lee:2010ik}
  R.~N.~Lee and V.~A.~Smirnov,
  JHEP {\bf 1102} (2011) 102
%
  [arXiv:1010.1334 [hep-ph]].

\bibitem{Gehrmann:2010ue}
  T.~Gehrmann, E.~W.~N.~Glover, T.~Huber, N.~Ikizlerli and C.~Studerus,
  JHEP {\bf 1006} (2010) 094
%
  [arXiv:1004.3653 [hep-ph]].

\bibitem{Gehrmann:2010tu}
  T.~Gehrmann, E.~W.~N.~Glover, T.~Huber, N.~Ikizlerli and C.~Studerus,
  JHEP {\bf 1011} (2010) 102
  doi:10.1007/JHEP11(2010)102
  [arXiv:1010.4478 [hep-ph]].

\bibitem{Chetyrkin:1993rv}
  K.~G.~Chetyrkin,
  In ``Oberammergau 1993, New computing techniques in physics research III'',
  559-563, [hep-ph/0212040].

\bibitem{Smirnov:2019qkx}
  A.~V.~Smirnov and F.~S.~Chuharev,
  arXiv:1901.07808 [hep-ph].

\bibitem{Lee:2013mka}
  R.~N.~Lee,
  J.\ Phys.\ Conf.\ Ser.\  {\bf 523} (2014) 012059
%
  [arXiv:1310.1145 [hep-ph]].

\bibitem{Ruijl:2017dtg}
  B.~Ruijl, T.~Ueda and J.~Vermaseren,
  arXiv:1707.06453 [hep-ph].

\bibitem{Chetyrkin:1997un}
  K.~G.~Chetyrkin, B.~A.~Kniehl and M.~Steinhauser,
  Nucl.\ Phys.\ B {\bf 510} (1998) 61
%
  [hep-ph/9708255].

\bibitem{Chetyrkin:2005ia}
  K.~G.~Chetyrkin, J.~H.~K\"uhn and C.~Sturm,
  Nucl.\ Phys.\ B {\bf 744} (2006) 121
%
  [hep-ph/0512060].

\bibitem{Schroder:2005hy}
  Y.~Schroder and M.~Steinhauser,
  JHEP {\bf 0601} (2006) 051
%
  [hep-ph/0512058].

\bibitem{Gerlach:2018hen}
  M.~Gerlach, F.~Herren and M.~Steinhauser,
  JHEP {\bf 1811} (2018) 141
%
  [arXiv:1809.06787 [hep-ph]].

\bibitem{Becher:2009cu}
  T.~Becher and M.~Neubert,
  Phys.\ Rev.\ Lett.\  {\bf 102} (2009) 162001
   Erratum: [Phys.\ Rev.\ Lett.\  {\bf 111} (2013) no.19,  199905]
%
  [arXiv:0901.0722 [hep-ph]].

\bibitem{progdata}
\verb|https://www.ttp.kit.edu/preprints/2019/ttp19-039/|.


\end{thebibliography}
\end{document}